\documentclass{aa}
\usepackage{graphicx}

\newcommand {\kms}{\mbox{km\,s$^{-1}$}}

\newcommand{\hcoplus}{\mbox{HCO$^+$}}

\begin{document}

\thesaurus{ 3(
		09.03.1; 
		09.07.1; 
		09.19.1; 
		10.07.1; 
		10.08.1; 
		10.19.3)  
             } 
\title{Search for molecular gas in HVCs by \hcoplus~ absorption}
   
\author{F. Combes\inst{1} and V. Charmandaris\inst{2}} 
\offprints{F.~Combes \hfill\break(e-mail: bottaro@obspm.fr)} 
\institute{
	Observatoire de Paris, DEMIRM, 61 Av. de l'Observatoire, 
	F-75014, Paris, France 
\and
Cornell University, Astronomy Department,
106 Space Sciences Bldg., Ithaca, NY 14853, USA
	} 
\date{Received 17 December 1999 / Accepted February 2000} 
\authorrunning{Combes \& Charmandaris} 
\titlerunning{Molecular absorption in HVCs}
\maketitle 
 
\begin{abstract}   
High-Velocity Clouds (HVCs) have radial velocities that cannot be
explained by the global Galactic rotation; their distances remain
mostly unknown, and their true nature and origin are still a
mystery. Some of them could be of galactic origin, or belong to tidal
streams drawn by the Milky-Way/ Magellanic Clouds interaction, or
could even be intergalactic clouds infalling onto the Local Group. In
the latter hypothesis, they play a major role in the hierarchical
formation scenario of the Milky-Way and are connected to the
Lyman-limit absorption systems. In any case, the determination of
their physical state (density, temperature, internal structure,
abundances, excitation) will help to discriminate between current
theories on their origin and nature. A recent UV measurement (Richter
et al 1999) has discovered for the first time in a HVC the molecular phase
that was previously searched for, without success, through CO
emission. Previous non detections could be due either to metallicity
problems, or insufficient excitation (because of low
density). Low-excitation molecular gas may, however, be detectable
though absorption. Here we report on a sensitive search for
\hcoplus(1-0) absorption lines in front of 27 quasars, already known
to be strong millimetric continuum sources. Except for one tentative
case, no detection was obtained in most HVCs, although \hcoplus(1-0)
was clearly detected towards galactic low-velocity clouds. We discuss
the implications of this result.  

\keywords{ ISM: clouds -- 
	ISM: general -- 
	ISM: structure -- 
	Galaxy: general -- 
	Galaxy: halo --
	Galaxy: structure }
\end{abstract} 
  
\section{Introduction} 

Since their discovery by Muller, Oort, \& Raimond (1963), HI clouds
moving with velocities that can not be explained by differential
Galactic rotation (often exceeding it by 100 \kms) have been the
target of numerous surveys.  Many of these high-velocity clouds (HVCs)
are located at intermediate and high Galactic latitudes (Giovanelli,
Verschuur, \& Cram 1973; Mathewson, Schwarz, \& Murray 1977; Wakker \&
van Woerden 1991) and do not appear to have any connection with the
gas in the Galactic disk. HVCs are often found in large HI complexes
with angular sizes 10--90$^\circ$. They cover from 10 to 37\% of the
sky, depending on the sensitivity of the studies (Murphy et al 1995).

The origin of HVCs is still unclear, mainly because the distances to
the individual complexes are in most cases unknown.  They could be
cold gas corresponding to the return flow in a Galactic fountain
(e.g. Houck \& Bregman 1990), or gas left over from the formation of
the Galaxy. Some HVCs are probably belonging to the tidal gas streams
torn from the Magellanic Clouds by the Milky-Way (Mathewson et al
1974, Putman \& Gibson, 1999).

Several authors have already explored the possibility that HVCs are
infalling primordial gas and have associated them with the Local Group
(see Wakker \& van Woerden 1997 for a thorough review). Recently Blitz
et al. (1999) re-examined this hypothesis, and simulating the
dynamical evolution of the Local Group galaxies, used the up-to-date
HI maps of HVCs to show that the HVCs are consistent with a dynamical
model of infall of the ISM onto the Local group. As such, they would
represent the building blocks of our galaxies in the Local group and
provide fuel for star formation in the disk of the Milky Way.  In
their model, HVCs contain altogether 10$^{11}$ M$_\odot$ of neutral
gas.  From their stability analysis, they conclude that there is
roughly 10 times more dark matter than luminous gas within each HVC,
and that these could correspond to the mini-halos which are able to
accumulate baryons, and can gather into filaments (e.g. Bond et al
1988, Babul \& Rees 1992, Kepner et al 1997). HVCs would therefore be
related to the hierarchical structure of the Universe (see e.g. Katz et
al 1996), and to the gas seen in absorption towards quasars
(Lyman-$\alpha$ forest and Lyman-limit lines). However, Giovanelli
(1981) has pointed out that the velocity distribution of HVCs does
not match that of the Local Group, but does favor an association with
the Magellanic Stream, the most obvious tidal feature of the interaction 
between the Galaxy and the Magellanic Clouds. The discrepancy
with the results from Blitz et al. (1999) comes from the
fact that the latter authors have not considered all observed HVCs,
but only a selection of them.

Maps of the brightest HVC complexes have revealed the existence of
unresolved structure at 10 arcmin resolution, which was further
resolved into high-density cloud cores at 1 arcmin resolution
(Giovanelli \& Haynes 1977; Wakker \& Schwarz 1991; Wakker \& van
Woerden 1991).  More generally, HVCs follow the fractal structure
observed in the whole interstellar medium (Vogelaar \& Wakker,
1994). The HI column densities in these cores are estimated to be
several times 10$^{20}$ cm$^{-2}$ and their temperatures are generally
between 30 and 300 K.  The central densities of individual clouds can
reach $>$ 80 cm$^{-3}$ D$_{kpc}^{-1}$, where D$_{kpc}$ is their
distance in kpc. Depending on the actual distance, which still remains
poorly determined, those conditions make the HVC cores possible sites
of star formation. HVCs have in fact been considered good candidates
for the source of young Population I stars at large distances from the
Galactic plane (see e.g. Sasselov 1993).

Attempts to measure the spin temperature of atomic hydrogen through
21cm absorption in front of background continuum sources have often
only resulted in upper limits (Colgan et al. 1990, Mebold et
al. 1991). A few detections have been reported (Payne et al. 1980,
Wakker et al. 1991, Akeson \& Blitz, 1999), with inferred spin
temperature as low as 36\,K, but in general most HVCs must have spin
temperatures larger than 200\,K or be very clumpy.

Most of our current knowledge of the gaseous content of HVCs comes
from HI observations. Efforts to search for molecular hydrogen using
CO emission lines have been so far unsuccessful (Wakker et al 1997),
since the sub-solar metallicity and/or low density of HVCs makes
direct CO emission line detection very difficult. However, optical
absorption lines have shown that HVCs are not completely devoid of
heavy elements (e.g. Robertson et al 1991, Lu et al 1994, Keenan et al
1995, Wakker \& van Woerden 1997).  The lines detected are from SiII,
CII, FeII, or CIV, but the strongest are from MgII (Savage et
al. 1993, Bowen \& Blades 1993, Sembach et al. 1995, 1998).
Metallicity studies have been done in HVCs to determine their
origin. If HVCs result from the Galactic fountain effect, their
metallicity should be solar while if they were associated with the
Magellanic or intergalactic Clouds, it could be even less than 0.1
solar. In fact, the determined abundances are around 0.1 solar, but
with much uncertainties, because of saturated lines, or dust depletion
(Sembach \& Savage 1996).  This average metallicity is compatible with
a Local-Group infall model, since X-ray observations have revealed
abundances of 0.1 solar in poor groups (Davis et al 1996), and even
0.3 solar in intra-cluster gas (Renzini 1997).

If HVCs were local analogues of Lyman-limit absorbing clouds,
background QSOs could enable us to detect molecules in absorption, a
task which is considerably easier to achieve.  Absorption has recently
been reconfirmed as a very powerful tool in the millimetric range
(Lucas \& Liszt 1996 hereafter LL96, Combes \& Wiklind 1996). A mm
molecular detection would advance our knowledge of HVCs, their
physical conditions, and their possible origin.  A first attempt has
been made in this domain by Akeson \& Blitz (1999) with only negative
results, using the BIMA and OVRO interferometers.  

Here we report on \hcoplus(1-0) absorption line observations,
made in the southern sky with the single dish 
15m ESO-SEST telescope, and
in the northern sky with the IRAM interferometer. The choice of
\hcoplus~  is justified because, due to its large dipole moment, it
is generally not excited in diffuse media (the critical density for
excitation is $\sim$ 10$^7$ cm$^{-3}$); therefore confusion with
emission is not a problem, contrary to the CO(1-0) line which requires
an interferometer to resolve out the emission. Furthermore, the
\hcoplus~ absorption survey carried out by Lucas \& Liszt (1994, 1996)
has revealed more and wider absorption lines than in CO, suggesting
that it might be a better tracer. In that study, the derived abundance
\hcoplus/H$_2$ was surprisingly large, of the order of 6$\times$10$^{-9}$,
and sometimes even an order of magnitude larger.  The details of the
observations are presented in Section 2.  Section 3 summarizes the
results, which are then discussed in Section 4.

\begin{figure*}[ht]
\includegraphics[width=\hsize]{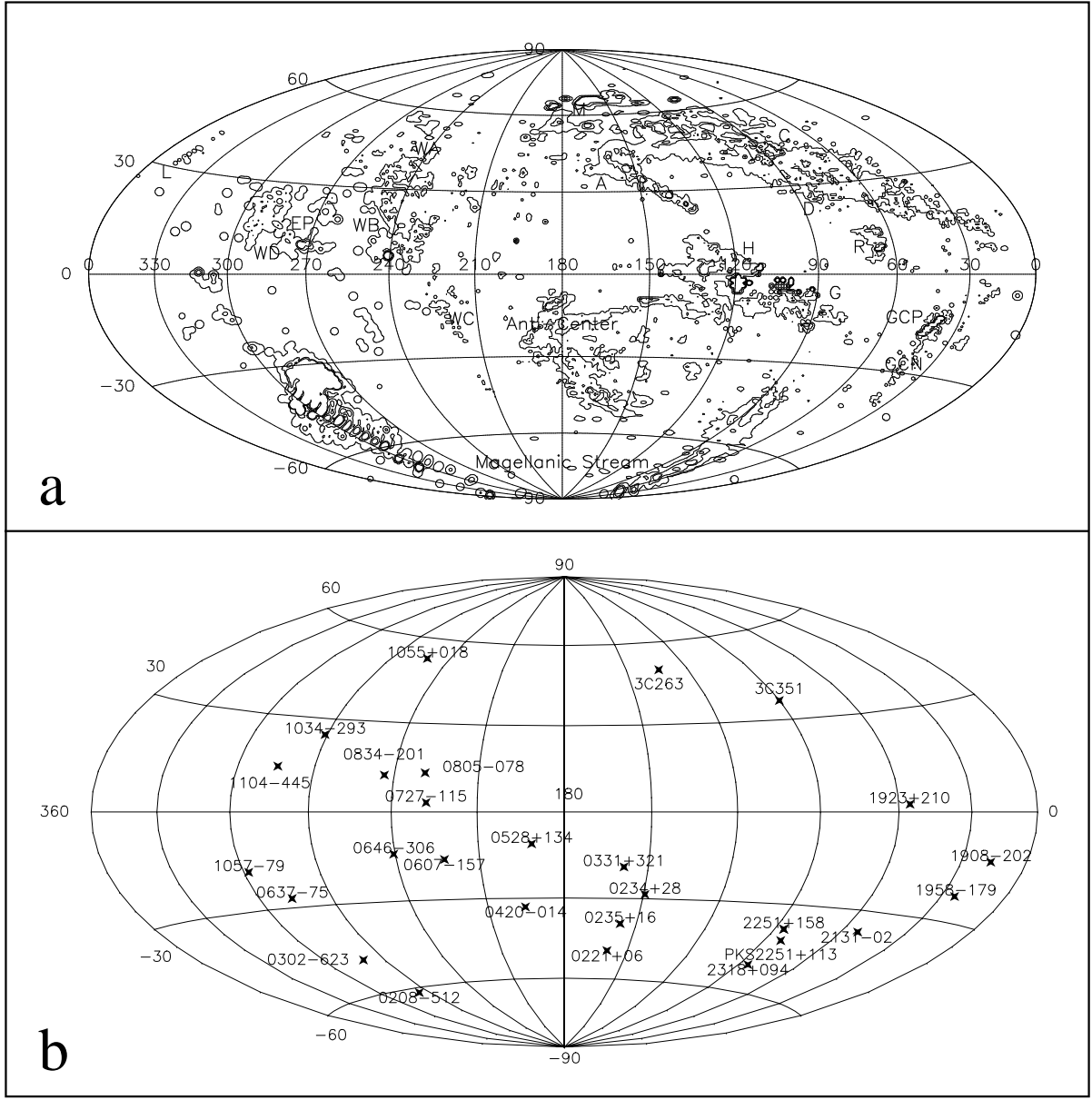}
\caption{ a) All-sky map of the high-velocity clouds, in HI 21cm line,
in Aitoff projection, from Wakker \& van Woerden (1997).  b) Location in the same
projection of the quasars observed in this work.}
\label{map}
\end{figure*}

\section{ Observations}

The observations in the southern sky 
were carried out with the 15m antenna of ESO-SEST in
La Silla (Chile), during 6-10 November 1999. We used two SIS receivers
at 3\,mm and 2\,mm simultaneously, to observe the \hcoplus(1-0) and
CS(3-2) lines at 89.188523 and 146.969033 GHz respectively.  Although
CS is less abundant, it would have been interesting to have both
molecules in case of detection.  The HPBW were 57'' and 34'' and the
main-beam efficiencies were $\eta_{\rm mb}=T_{\rm A}^*/T_{\rm
mb}$=0.75 and 0.66 respectively at the two mentioned frequencies.  The
backends were accousto-optic spectrometers (AOS) both at
high-resolution (HRS) and low-resolution (LRS). The corresponding
channel spacings (velocity resolutions) were 0.144 (0.268) and 0.087
(0.163) \kms~ at \hcoplus~ and CS for HRS and 2.32 (4.7) and 1.4 (2.8)
\kms~ respectively for LRS.  The number of channels were 1000 and 1440
for high and low resolution backends respectively, so that the
velocity coverage was 140 and 3352\,\kms~ at high and low resolution for
the \hcoplus~ line, and 85 and 2057\,\kms~ for CS respectively.  Since
all backends were centered of the expected HVC velocity, there was no
problem detecting any galactic line around V=0\,\kms~ with the
low-resolution, but it was most of the time outside of the range at
high-resolution. We have retrieved the \hcoplus~ absorptions already
reported by Lucas \& Liszt (1996) with the IRAM interferometer. In
cases where dilution in large velocity channels made detection
problematic, we shifted the high-resolution backend to zero velocity
to verify the detection.  Pointing was corrected regularly (every 2
hours) using known SiO masers (and the 3\,mm receiver retuned
accordingly). The weather was clear throughout the run, and the
typical system temperatures were 180\,K for both the 3 and 2\,mm
ranges. The observing procedure was dual beam switching at high
frequency (6\,Hz) between same elevation positions in the sky (beam
throw of 2' 27''), to eliminate atmospheric variations.  Each source
was observed on average for 2 hours, reaching about 3.5\,mK of rms
noise in channels of 1.4\,MHz (4.7\,\kms~ at \hcoplus).  

Because the width of the absorption lines may be expected to be small
(even though the \hcoplus~ absorption lines are the wider mm lines, and
therefore more favorable for detection), a concern should be noted
that line profiles may be diluted in the low resolution spectrograph.
The technique proved to be valuable, allowing us to retrieve the
\hcoplus(1-0) absorptions already detected by LL96 for galactic
clouds, near zero velocity (see below).  We also wished to check that
the \hcoplus(1-0) absorption was not hampered by emission. In one of
the sources (2251+158) we observed an offset position (with a beam
throw of 12' for the beam-switch) to check for emission. Emission was
not detected at the same signal-to-noise at which the absorption was
clearly present.

The observations in the northern sky 
were carried out with the IRAM interferometer in Plateau de Bure
(France), during July, October and November 1999. 
The interferometer data were made with the standard D configuration
(see Guilloteau et al. 1992). The array comprised 4 15--m telescopes.
The receivers were 3--mm SIS, giving a typical system temperature
of 150--200\,K. One of the source, 3C 454.3,  was used itself as a phase reference,
while bandpass and amplitude calibrations were done using also sources
such as 3C 273, MWC 349, 1823+568, 2145+067. 
The data reduction was made with the standard CLIC software.
The synthesized beams are of the order of 6$\times$8 arcsec. 
 The auto-correlator was used in three overlapping configurations
of bandwidth 160, 80 and 40 MHz and a respective resolution of 
2.5, 0.625 and 0.156 MHz, giving a maximum velocity resolution of
0.5 \kms. The largest bandwidth observed is 540 \kms.

\begin{table*}
\caption[]{List of objects observed at SEST, and their detected 3 and 2\,mm continuum.\\
The 4 objects in the secondary list have been observed with IRAM interferometer.}
\begin{tabular}{lcccccccc}
\\
\hline
\\
{QSO}  &    {R.A.} & {Dec.} & Long & Lat  & {V$_{lsr}$} & HVC  &  F$_{3mm}$ & F$_{2mm}$ \\
          & (B1950) &(B1950)& (deg)& (deg)& \kms      & complex&  Jy           &   Jy     \\ 
\\
\hline
\\
\object{0208-512}  & 02:08:57.1 &   -51:15:08 & 276.10 & -61.78 &  180. & MS  & 4.96 & 4.41\\
\object{0221+06}   & 02:21:49.9  &  +06:45:50 & 159.72 & -49.12 & -230. & AC  & 0.83 & 0.29\\
\object{0234+28}   & 02:34:55.5  &  +28:35:12 & 149.46 & -28.53 & -140. & AC  & 3.42 & 2.06\\
\object{0235+16}   & 02:35:52.6  &  +16:24:04 & 156.77 & -39.11 & -230. & AC  & 2.01 & 1.41\\
\object{0302-623}  & 03:02:48.4  &  -62:23:06 & 280.23 & -48.67 &  200. & MS  & 0.57 & 0.53\\
\object{0331+321}  & 03:31:40.1  &  +32:08:37 & 158.70 & -18.99 & -140. & AC  & 0.12 & 0.06\\
\object{0420-014}  & 04:20:43.5  &  -01:27:28 & 195.29 & -33.14 & -180. & AC  & 2.36 & 1.76\\
\object{0528+134}  & 05:28:06.8  &  +13:29:42 & 191.37 & -11.01 & -180. & AC  & 2.83 & 1.67\\
\object{0607-157}  & 06:07:26.0  &  -15:42:03 & 222.61 & -16.18 &  110. & WC  & 4.96 & 2.35\\
\object{0637-75}   & 06:37:23.3  &  -75:13:34 & 286.37 & -27.16 &  250. & MS  & 2.12 & 1.47\\
\object{0646-306}  & 06:46:19.6  &  -30:40:54 & 240.54 & -14.11 &  190. & WC  & 0.76 & 0.35\\
\object{0727-115}  & 07:27:58.5  &  -11:34:53 & 227.77 &   3.14 &  125. & WB  & 1.30 & 0.88\\
\object{0805-078}  & 08:05:49.6  &  -07:42:23 & 229.04 &  13.16 &  125. & WB  &  1.42 & 0.88\\
\object{0834-201}  & 08:34:24.6  &  -20:06:30 & 243.57 &  12.23 &  150. & WB  & 0.71 & 0.44\\
\object{1034-293}  & 10:34:55.9  &  -29:18:28 & 270.95 &  24.84 &  120. & EP  & 0.71 & 0.20\\
\object{1055+018}  & 10:55:55.3  &  +01:50:04 & 251.51 &  52.77 &  110. & WA  & 1.89 & 1.32\\
\object{1057-79}   & 10:57:50.3  &  -79:47:39 & 298.01 & -18.28 &  250. & EP  & 1.18 & -- \\
\object{1104-445}  & 11:04:50.4  &  -44:32:53 & 284.17 &  14.22 &  150. & EP  & 1.30 & 0.68\\
\object{1908-202}  & 19:08:12.5  &  -20:11:55 &  16.87 & -13.22 &  1 & 2.36 & 1.76\\
\object{1923+210}  & 19:23:49.8  &  +21:00:23 &  55.56 &   2.26 & -200. & R?$^{\mathrm{a}}$   & 2.36 & 2.00\\
\object{1958-179}  & 19:58:04.6  &  -17:57:16 &  24.01 & -23.11 &  234. & GCP & 1.18 & 0.65\\
\object{2131-02}   & 21:31:35.3  &  -02:06:41 &  52.39 & -36.50 & -206. & GCN & 1.06 & 0.59\\
\object{2251+158}  & 22:51:29.5  &  +15:52:54 &  86.11 & -38.18 & -300. & MS  & 6.14 & 4.12\\
\object{2318+094}  & 23:18:12.1  &  +04:57:23 &  85.42 & -50.92 & -350. & MS  & 0.47 & 0.29\\
\\
\hline
\\
\object{3C263}         &  11:37:09.3  &  +66:04:27 & 134.16 &  49.74 & -185.& C  & 0.06 & -- \\ 
\object{3C351}         &  17:04:03.5  &  +60:48:31 & 90.08 &  36.38 & -180.& C  & 0.02 & -- \\ 
\object{3C454.3}       &  22:51:29.5  &  +15:52:54 & 86.11 &  -38.18 & -397.& MS  & 5.44 & -- \\ 
\object{PKS2251+113}&  22:51:40.5  &  +11:20:39 & 82.78 &  -41.94 & -374.& MS  & 0.02 & -- \\ 
\hline
\end{tabular}

\begin{list}{}{}
\item[$^{\mathrm{a}}$] might belong instead to Cloud 274 or 283
(B. Wakker, priv. comm.) 
\end{list}  

\label{tab1}
\end{table*}

\begin{figure}[ht]
\includegraphics[width=\hsize]{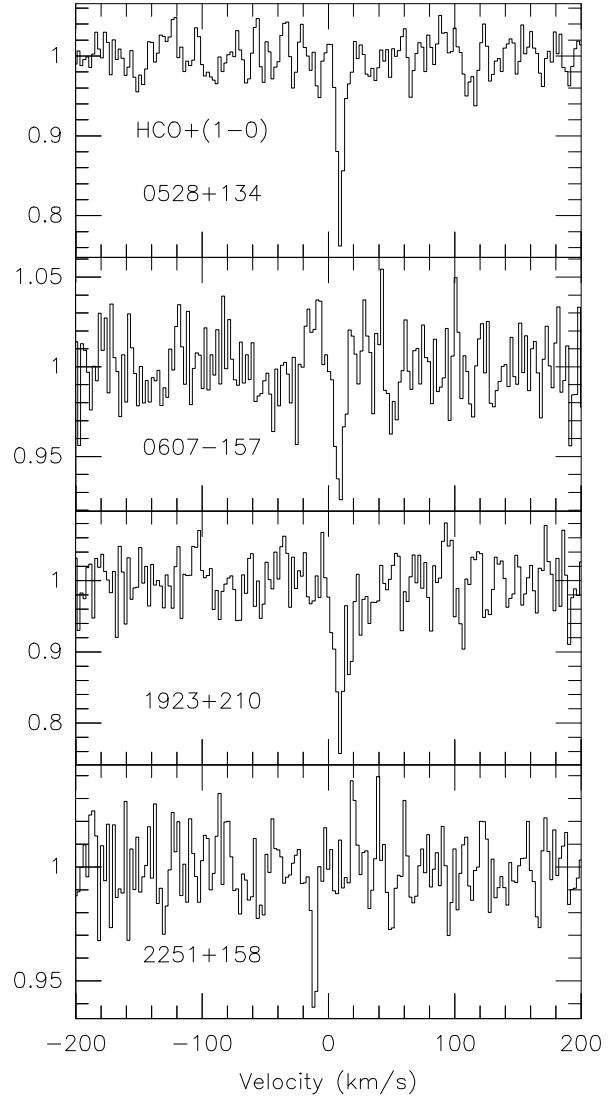}
\caption{ Galactic clouds seen in absorption in the \hcoplus(1-0) line in front
of the observed sources. One of the absorption lines is new
(1923+210).  The lines are very narrow, and are diluted in velocity,
therefore appear of lower intensity than previously reported
(LL96). The channel spacing is 2.3\,\kms~ and the velocity resolution is
4.7\,\kms. The spectra are normalized to the total continuum
detected. }
\label{spec}
\end{figure}

\section{Results}

Our sources were selected from the strong millimeter quasars that were
seen in projection over one of the 17 HVC complexes identified, in the
Wakker \& van Woerden (1991) recapitulating map. Table \ref{tab1}
displays the characteristics of the sources observed, the name and
expected velocity of the HVCs along each line of sight, following the
notation of the HVC catalogue of Wakker and van Woerden (1991), as
well as the measured continuum flux in Jy at 3 and 2\,mm, for all 24
sources observed at SEST at the date of observation (6--10 November 1999).  
The 4 sources observed with the IRAM interferometer (also in
Table \ref{tab1}), have been selected to have MgII detected
in absorption by Savage et al. (1993), at very high velocities. The HI 
in emission at these high velocities was measured by the NRAO-43m
HI survey towards 143 quasars of Lockman \& Savage (1995). The detection
of MgII insures the presence of a minimum metallicity towards these
lines of sight (larger than 0.32 solar, in 3C 454.3 for instance), 
favorable for the detection of \hcoplus. 
The positions of the observed sources are plotted in Figure 1 vis-a-vis
an HI map of HVCs in Aitoff projection. Note that one source has been
observed both with IRAM and SEST (3C 454.3).

The spectra were then normalized to the continuum flux, and the upper
limit at 3$\sigma$ of the optical depth $\tau$ in absorption was
computed in each case assuming that the surface filling factor is $f$=1,
i.e., that the absorbing molecular material covers completely
the extent of the mm continuum source, in projection.
If $T_{cont}$ is the observed continuum antenna temperature
of the background source, and $T_{abs}$ the amplitude in temperature
of the absorption signal, then the optical depth is:

$$
\tau = - ln ( 1 - \frac{T_{abs}}{f T_{cont}} )
$$ 
The 3$\sigma$ upper limits of $\frac{T_{abs}}{T_{cont}}$ as
measured in 0.56\,\kms~ channels is listed in Table \ref{tab2}.

The total column density of \hcoplus, observed in absorption between
the levels $u$ $<$--- $l$ with an optical depth $\tau$ at the center of
an observed line of width $\Delta v$ at half-power, is:

$$ N_{\hcoplus} = {{8\pi}\over{c^3}} f(T_x) {{\nu^3 \int\tau dv }
\over {g_u A_u} }, $$ where $\nu$ is the frequency of the transition,
$g_u$ the statistical weight of the upper level (= 2 J$_u$+1), $A_u$
the Einstein coefficient of the transition (here $A_u$ =
3$\times$10$^{-5}$ s$^{-1}$), $T_x$ the excitation temperature, and 
$$
f(T_x) = {{Q(T_x) exp(E_l/kT_x)} \over { 1 - exp(-h\nu/kT_x)}} 
$$
where $Q(T_x)$ is the partition function.  We assumed statistical
equilibrium an excitation temperature, close to the cosmic background
temperature, i.e. $T_x$ = 3\,K, because of the large critical density
needed to excite \hcoplus. We also expect a very narrow linewidth,
comparable to the detected lines, and adopted $dv$ = 1.1\,\kms~ to
derive the upper limits in Table \ref{tab2}. The \hcoplus/H$_2$ abundance
was conservatively taken as 6$\times$10$^{-9}$, but it must be kept in
mind that it could be more than an order of magnitude higher
(e.g. Lucas \& Liszt 1994, Wiklind \& Combes 1997), and therefore the
derived H$_2$ column densities could be correspondingly lower.
However, if the metallicity is lower than solar, the corrections
should go in the reverse sense.

An indicative HI column density of the high-velocity gas was also
estimated using the HI surveys of Stark et al (1992) and Hartmann \&
Burton (1997) for the northern hemisphere, and Bajaja et al. (1985) for
our 5 southernmost sources. This column density is rather uncertain
since it has been smoothed over large areas (at least one half of a
degree) while the material that could appear in absorption extends
over milli-arcseconds, similar to the background millimeter
sources. At small scales the HI column density could be much higher
than the values presented.

Figure \ref{spec} shows some of the low velocity detections. All of
them have already been discovered by Lucas \& Liszt (1996), except the
one at 1923+210 which is new. In the low-resolution backends, though,
they are barely resolved and have somewhat reduced peak intensity. To
check this, we have re-tuned to their central velocity, and centered
the high-resolution spectrograph at V$\sim$ 0. In the HRS AOS, the
absorptions had indeed stronger intensities and were in all cases
compatible with previously reported values (the integration time
though, for the retuned spectra, was not enough to obtain a high S/N).

\begin{figure}[ht]
\includegraphics[width=\hsize]{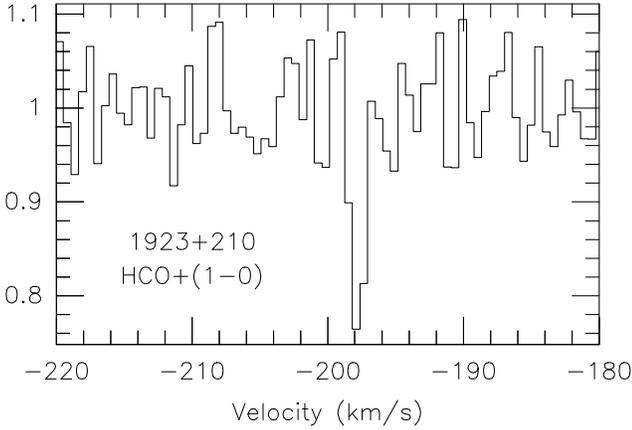}
\caption{ Tentative detection of \hcoplus(1-0) absorption in the HVC in front
of 1923+210 (at 5.5$\sigma$).  The channel spacing is 0.56 \kms~ and
the velocity resolution is 1.1\,\kms. The spectrum is normalized to the
total continuum detected (2.36 Jy). }
\label{tenta}
\end{figure}

Towards the source 1923+210, a tentative detection at the expected HVC
velocity was obtained, as shown in Figure \ref{tenta}. It is very
narrow, and was detected clearly only in the HRS back-end.

Akeson and Blitz (1999) have recently reported upper limits in CO
absorption with the BIMA and OVRO interferometers towards 7 continuum
sources. We have no sources in common, and therefore we increase the
statistical significance of the upper limits.  They also searched for
HI absorption in HVCs with the VLA, and have positive results only in
the gas associated to the outer arm of our Galaxy. They conclude that
true HVCs are very weak HI and molecular absorbers.  HI absorption in
HVCs has been searched with single dish or interferometer several
times, without much success (Payne et al. 1980, Colgan et al. 1990,
Wakker et al. 1991). The fact that only $\sim$ 5\% of the lines of
sight towards HVCs show HI absorptions, while this frequency reaches
100\% for normal galactic gas (Dickey et al. 1983), sheds some light
on the physical structure of the clouds.


\begin{table}
\begin{flushleft}
\caption[]{Derived column densities (upper limits assuming 1.1\,\kms~ linewidth)}
\begin{tabular}{llcccc}
\\
\hline
{QSO}  & $\frac{T_{abs}}{T_{cont}}$  &    
N(\hcoplus)          & N(H$_2$)           &  N(HI)    & $\frac{H_2}{HI}$  \\
       & 3\,$\sigma\,^{\mathrm{a}}$  &  
  10$^{11}$  & 10$^{19}$ &  10$^{19}$ &  3 $\sigma$ \\ 
       &  &  cm$^{-2}$   & cm$^{-2}$ &  cm$^{-2}$ &  \\ 
\hline
\\
 0208-512  &   0.08    &  1.2  &  2.0   &  1.4   &  1.4  \\
 0221+06   &   0.67    & 17.0  & 28.3   &  1.6   & 17.7  \\
 0234+28   &   0.12    &  1.9  &  3.2   &  2.2   &  1.4  \\
 0235+16   &   0.22    &  3.7  &  6.2   &  1.6   &  3.9  \\
 0302-623  &   0.53    & 11.0  & 18.3   &  2.3   &  7.9  \\
 0331+321  &   $>$1    & --    &  --    &  0.50  &   --  \\    
 0420-014  &   0.24    &  4.1  &  6.8   &  0.15  & 45.3  \\
 0528+134  &   0.11    &  1.7  &  2.8   &  1.1   &  2.5  \\
 0607-157  &   0.08    &  1.2  &  2.0   &  2.0   &  1.0  \\
 0637-75   &   0.27    &  4.7  &  7.8   &  6.4   &  1.2  \\
 0646-306  &   0.53    & 11.0  & 18.3   &  2.0   &  9.1  \\
 0727-115  &   0.40    &  7.6  & 12.7   &  3.0   &  4.2  \\
 0805-078  &   0.40    &  7.6  & 12.7   &  4.7   &  2.7  \\
 0834-201  &   0.53    & 11.0  & 18.3   &  7.3   &  2.5  \\
 1034-293  &   0.92    & 38.0  & 63.3   &  4.6   & 13.8 \\
 1055+018  &   0.27    &  4.7  &  7.8   &  3.6   &  2.2  \\
 1057-79   &   0.80    & 24.0  & 40.0   &  3.9   &  10.2  \\
 1104-445  &   0.24    &  4.1  &  6.8   &  4.3   &  1.6  \\
 1908-202  &   0.20    &  3.3  &  5.5   &  3.0   &  1.8  \\
 1958-179  &   0.40    &  7.6  & 12.7   &  3.0   &  4.2  \\
 2131-02   &   0.40    &  7.6  & 12.7   &  0.18  & 70.5  \\
 2251+158  &   0.04    &  0.61 &  1.0  &  0.22  &  4.5  \\
 2318+094  &   0.92    & 38.0  & 63.3   &  2.2   & 28.8  \\
\\
\hline
\\
 3C263      &   0.5    & 10.4  & 17.4   &  0.2   & 87.  \\
 3C351      &   0.9    & 35.0  & 58.4   &  0.9   & 65.  \\
 3C454.3  &   0.007    & 0.10  & 0.17   &  0.12   & 1.4  \\
 PKS2251+113 &   0.9    & 35.0  & 58.4   &  0.48   & 121.  \\
\hline
\end{tabular}

\vspace*{0.5cm}
 Tentative  Detection  \\
 (with a linewidth precisely 1.1\,\kms)\\
\begin{tabular}{llcccc}
{QSO}  & $\frac{T_{abs}}{T_{cont}}$  &    N(\hcoplus)    & N(H$_2$)    &  N(HI)    & $\frac{H_2}{HI}$  \\
       &  &  10$^{11}$  & 10$^{19}$ &  10$^{19}$ &  3 $\sigma$ \\ 
       &  &  cm$^{-2}$   & cm$^{-2}$ &  cm$^{-2}$ &  \\ 
\hline
\\
 1923+210  &   0.25         & 4.3  &  7.2    &  1.  & 7.2  \\
\hline
\end{tabular}

\begin{list}{}{}
\item[$^{\mathrm{a}}$] in 0.56 \kms~ channels
\end{list}

\label{tab2}
\end{flushleft}
\end{table}

\section{Discussion and Conclusion}

\subsection{ The diverse nature of HVCs}

It is still difficult to tackle the problem of the origin of HVCs,
since they may be diverse in nature, with possibly different formation
histories. The first idea that they are clouds infalling towards the
Milky Way (Oort, 1966) was proposed since only negative velocities
clouds were discovered at this epoch. However, since differential
galactic rotation makes an important contribution to those velocities
and the tangential velocity of the HVCs is also unknown, it is
difficult to ascertain their true three dimensional motion. Now that
almost an equal number of clouds are detected with positive high
velocities, and that is also true in the various reference frames
 (e.g. Wakker 1991), it is believed there must be other
explanations for at least some of them.  Some HVC complexes have been
proven to be associated with the Galaxy, in being very close (less
than 10 kpc) due to absorption detected in front of Galactic halo
stars (e.g. Danly et al. 1993, van Woerden et al. 1999).  This
supported a Galactic origin for some of the HVCs and more specifically
the Galactic fountain model, where gas is ejected in the halo by the
star formation feedback mechanisms.  The main problems with this
scenario are the low metallicity observed in most lines of sight
(even though it could explain why clouds have nearly solar metallicity,
e.g. Richter et al 1999), and the extreme velocities sometimes reached
(up to -450\,\kms~ in the negative side).  
Alternatively, HVCs could be transient structures,
now becoming bound to the Galaxy: an ensemble or complex of HVCs has
been identified as tidal debris from the Magellanic Clouds (Mathewson
et al. 1974), and is called the Magellanic Stream. Few of them could
come from tidal debris from interactions with other dwarf galaxies but
the corresponding dwarf galaxies either have not been identified yet,
or they have been already dispersed (e.g. Mirabel \& Cohen 1979).

A key point is to consider whether HVCs are transient structures or
coherent and self-gravitating. With only their column density observed
in HI, to be self-gravitating they should be at least at 1\,Mpc
distance, and in average at 10 Mpc, therefore outside the Local Group
(Oort, 1966). This distance can be reduced by a factor 10, if the
total mass of HVC is taken to be 10 times their HI mass, composed mostly 
of dark matter (Blitz et al. 1999). Since the specific kinetic energy is
in M/R $\propto$ Distance, then the clouds would have to be only at about 1\,Mpc, and would consequently be part of the Local Group.  The main difficulty with this
model of mini-haloes merging with our Galaxy, is that other
high-velocity clouds should have been observed around or in between
external galaxies, which is not the case (Giovanelli 1981, Zwaan et
al. 1997, Banks et al. 1999).  Moreover, the derived typical
dimensions and mass of these systems do not correspond to any observed
object or any extrapolation of mass ranges (they have dwarf masses,
but are very extended in size). 

Another possibility, however, could be that HVCs represent gas left
over from the formation of the Galaxy: they form a system out of
equilibrium, but bound to the Galaxy, and are now raining down to the
Galactic disk. This would explain several observed characteristics, such as the
4-10 kpc distance determined for some complexes, and the low
metallicity of many of them (e.g. Wakker et al. 1999).

\begin{figure}[ht]
\includegraphics[width=\hsize]{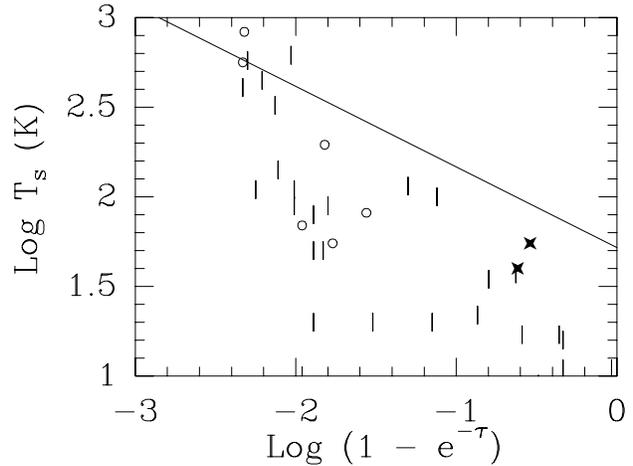}

\caption{ Spin temperature T$_s$ of the HI in HVCs derived from absorption/emission
measurements, as a function of the optical depth $\tau$ of the absorption.
Positive detections are from Payne et al. 1980 (filled stars) and Wakker
et al. 1991 (unfilled circles). Upper limits in HI absorption lead to
lower limits in T$_s$ and are plotted by vertical bars (Payne et al. 1980,
Wakker et al. 1991, Akeson \& Blitz 1999). The line is that fitted to
the detections of low-velocity gas (Payne et al. 1983).  }
\label{ts}
\end{figure}

\subsection { How many molecular absorptions could be expected? }

Given the low HI column densities reported in Table \ref{tab2}, it could
seem hopeless to detect molecular absorption, with the present
sensitivities. But this is true only if the HVC gas is homogeneous.
 In fact, from previous HI observations, we expect that like the gas in the galactic disk,the HVC gas is also composed  of several components,
with clumps of much higher column densities, only detectable at higher
spatial resolution. This information can be deduced from 
existing  HI observations in emission and absorption, towards HVCs.
The optical thickness $\tau$ is related to the
column density N(HI) (cm$^{-2}$), the spin temperature T$_s$(K) and the
FWHM velocity width $\Delta V$ (\kms) by 
$$
\tau = \frac {N(HI)}{1.8\times10^{20}} \,  \left(\frac{50}{T_s}\right) \,
\left(\frac{20}{\Delta V}\right)
$$ 
Since the profile extends at least over 20\,\kms~ in emission, and the
spin temperature has been determined to be at least 50\,K, if there
was only one component, the derived optical depth would have
to be very low, $\tau < 0.1$. Absorption and emission studies in the
Galactic plane have shown though that there is indeed more than one 
component. If there was only one, both emission and absorption
profiles would look similar, which is not what is observed.
In fact, there are at least two components; a warm diffuse
inter-clump component with cloudlets which are cold, narrow in
velocity, and more optically thick (Garwood \& Dickey 1989). If the
surface filling factor of the two components are f$_1$ and f$_2$, for
spin temperatures T$_1$ and T$_2$, and optical depths $\tau_1$ and
$\tau_2$, the observed ratio between the absorption depth
$\frac{T_{abs}}{T_{cont}}$ and the antenna temperature of the emission
$T_{em}$ is $$
\frac{T_{abs}}{T_{cont} T_{em}} =
\frac{f_1 (1-e^{-\tau_1}) +f_2 (1-e^{-\tau_2}) } 
{T_1 f_1 (1-e^{-\tau_1}) + T_2 f_2 (1-e^{-\tau_2}) } 
$$
In this formula, it is obvious that each component is weighted
according to its mass if it is optically thin (i.e. weight $\propto f
\tau$), but considerably less if $\tau >> 1$. Therefore, if the cold
component is optically thick, the derived spin temperature will be
overestimated\footnote{In fact the mass of the cold component would be
itself under-estimated, since the absorption often saturates, while
the emission, dominated by the warm gas, saturates less}.  

Are HVC clouds of the same nature than normal low-velocity galactic
clouds, and have they similar  small-scale structure? For normal HI
clouds, Payne et al. (1983) have determined (from absorption/emission
comparisons) that the weighted 
average spin temperature is decreasing as the optical depth 
increases (cf line in  Figure \ref{ts}). There is only a small
scatter in this relation, which means that if the cold and thick
component is made of cloudlets, their size must be smaller than that
of their background continuum sources, i.e. a fraction of an arcmin, and their
number must be large accordingly (N $>$ 100). We do not see the
situation where absorption features are deep and rare (statistically
the T$_s$-$\tau$ relation will still hold, but with large
scatter). The data favor a model, in which the cloudlets are quite
small ($<$ 0.1 pc) and numerous, and they are weighted according to
their surface filling factor $f$ both in emission and in absorption for
any line of sight. The existence of such small scale structure is also
confirmed by VLBI HI absorption (Faison et al. 1998), where sizes down
to $\sim$ 20 AU are detected.  It is also very likely that the
structure, apart from these smallest fragments, has no particular
scale.  If it is a fractal, statistically, there is a correlation
between the optical depth between scales with a limited scatter. This
would explain the observed correlation between emission and absorption
at different scales.  When the emission/absorption measurements for
HVCs are considered (cf Figure \ref{ts}) the lower limits on derived
spin temperature are compatible with the detections on the normal
low-velocity gas and their T$_s$-$\tau$ relation. The spin temperature
of HVCs is a weighted mean of the warm and cold components, with the same
mixtures as the one observed in normal low-velocity clouds of the
galactic plane. The difficulty to find HI in absorption in HVCs
corresponds then only to their low average column density,
and not to a different physical structure.

As for molecular absorptions, the column densities to which a
detection is possible is even larger. The corresponding scales must be
smaller, and consequently the corresponding line-widths narrower.
This is observed for low-velocity absorptions, where the detected
line-widths are as narrow as 0.6\,\kms (LL96). The detected optical
depths are larger on average, and the probability to underestimate the
column density is higher, because of saturation. (One should note
though that the apparent optical depth is low because of spatial and
velocity dilutions.) The probability of detection of the \hcoplus(1-0)
absorption has been estimated to be 30\% as large as the 21cm
absorption for galactic clouds (LL96).  The presently observed low
probability to find an \hcoplus(1-0) absorption is therefore
expected. In addition, there could be a column density threshold for
self-shielding against photo-dissociation, that hampers molecular
observations.  This threshold has been estimated for CO emission to
N(H$_2$) $\sim$ 4$\times$10$^{20}$\,cm$^{-2}$ (or equivalently of the
order of 10$^{12}$\,cm$^{-2}$ for
\hcoplus). If our tentative detection is confirmed, given
the low metallicity of the HVCs, this will support the existence of
clumps of high column densities in this medium.
  The fact that molecular absorption is more frequent with respect
to emission than atomic absorption is related to the excitation 
mechanism. Large volumic critical densities are required 
to excite molecules above the cosmic background, and this
is in particularly true for \hcoplus. This makes absorption
techniques the more promising to probe the molecular content
of HVCs in the future.

\subsection { Physical nature and distance of the gas}

\begin{figure}[ht]
\includegraphics[width=\hsize]{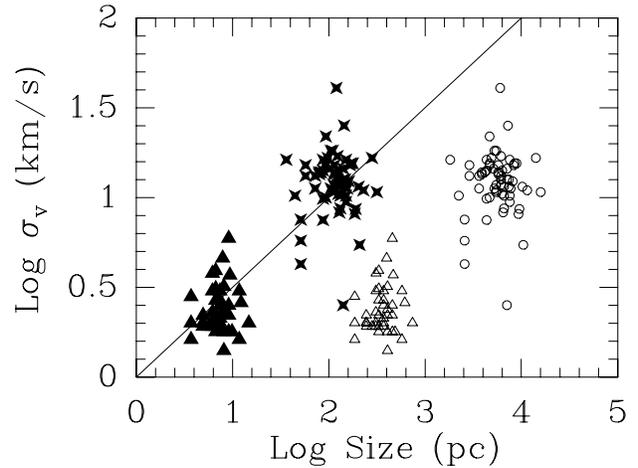}

\caption{ Size-linewidth plot for HVCs, observed with two
spatial resolutions. The first set is the compact isolated HVCs
from the Braun \& Burton (1999) catalog, with two assumed distances,
20 kpc (filled stars), and 1 Mpc (open circles). The second set is the 
clouds identified by Wakker \& Schwarz (1991) with the Westerbork 
interferometer, with the same two assumed distances, 
20 kpc (filled triangles), and 1 Mpc (open triangles). The line is the
relation derived for clouds in the Milky Way (Solomon et al. 1987).  }
\label{bb}
\end{figure}

 Since the HVCs appear to follow in projection the same fractal
properties as the normal low-velocity gas (Vogelaar \& Wakker 1994),
it could be interesting to develop an insight in their distance from
their size-linewidth relation. It is now well established that clouds in the
interstellar medium (either molecular or atomic) are distributed 
according to a self-similar hierarchical structure, 
characteristic of a fractal structure (Falgarone et al. 1991;
Stanimirovic et al. 1999; Westpfahl et al. 1999).  Such a scaling 
relation between mass and size, can also lead to a relation
between size and line-width, provided that the structures
are virialized. 
In particular, various clouds at all scales obey a power-law relation
between sizes $R$ and line-widths or velocity dispersion $\sigma$: $$
\sigma \propto R^q $$ with $q$ between 0.35 and 0.5 (e.g. Larson 1981,
Scalo 1985, Solomon et al 1987).  We have plotted this latter
relation, together with the sizes and line-widths of the 65 well
defined isolated HVCs, catalogued by Braun \& Burton (1999). This
catalog is expected to be free from galactic contamination as well as
from blending along the line of sight, because of the isolation criterium
imposed for their selection.  To compute their sizes, two distances
were assumed, either 20 kpc, or 1 Mpc, and the geometrical mean
between major and minor axis was calculated. At a distance of 20\,kpc,
the clouds fall on the relation corresponding to Giant Clouds in the
Galaxy. Note that the large scatter is due to the fact that this
choice of distance is only one order of magnitude, and the HVCs are
certainly not all at the same distance.  This set of clouds has been
determined with a spatial resolution of half a degree.  The clouds
determined at higher spatial resolution have correspondingly narrower
line-widths. In Figure \ref{bb}, we have also plotted the
characteristics of clouds determined with the Westerbork interferometer,
at 1 arcmin resolution (Wakker \& Schwarz 1991).  They also fit the
galactic clouds relation on average, if their distance is chosen to be
20 kpc, with large scatter since individual distances are also not
known. In fact, HI observations at different spatial resolutions 
emphasize a particular scale of the hierarchical structure (see e.g 
the 10' resolution observations by Giovanelli \& Haynes 1977). This
hierarchical structure is similar to what is observed for ``normal''
galactic clouds, in the sense that the same fraction 20-30\% of the
single dish flux is retrieved in the interferometer data (Wakker \&
Schwarz 1991).

The size-linewidth relation has been widely observed up to 100pc in
size, the largest size for self-gravitating clouds in the Galaxy, but
it might appear questionable to extend it to higher scales, where the
gas would not be self-gravitating.  At these scales, the gas is bound
into largest self-gravitating structures, including stars or dark
matter. However, even at these scales, the gas should trace the
gravitational potential of the bound structure it is embedded in,
and share the corresponding velocity
dispersion; such a relation is observed for instance in the form of
the Tully-Fisher relation in galaxies. The main point is that the gas
should reveal velocity profiles in emission that should grow wider
with the distance, if it belongs to an assumed self-gravitating remote
system. The observed profile width is therefore a distance indicator.

\subsection {Conclusion}

We have searched for \hcoplus(1-0) absorption towards 27 high velocity
clouds, in front of remote radio-loud quasars. The technique is
efficient, since we detect the existing absorption due to low velocity
galactic clouds, for our low latitude sources. Only one tentative HVC
detection is reported. If confirmed, this indicates the presence of
small scale cloudlets, at low excitation and high column densities,
prolonging the hierarchical structure already observed in the atomic
component at larger scale. When this hierarchical structure is
compared to the one observed for low-velocity galactic clouds, a good
fit is obtained for the self-similar relation between sizes and
line-widths, if the HVCs are on average at 20\,kpc distance.  Since mm
molecular absorptions are expected to be more frequent than emission
for these low column density HVCs, this absorption technique appears
promising to probe the molecular component of HVCs, already directly
detected by UV H$_2$ absorption lines (Richter et al 1999).

\begin{acknowledgements}
 We are very grateful to R. Giovanelli for useful
comments on the manuscript, to B. Wakker and F. Mirabel for
stimulating discussions, and an anonymous referee for constructive
criticism. We also thank the SEST staff for their kind help
during the observations, and Raphael Moreno and Robert Lucas
for their assistance in the IRAM interferometer data reduction.
\end{acknowledgements} 
%

\end{document}